# A Prototypical Decision-Support Tool for Household Energy Management: A New Zealand Case Study

**Abdollah Baghaei Daemei**[1*]
1- Building Performance Analysis Lab, Tech Innovation Experts Ltd., Auckland, New Zealand
Corresponding author: abdollah.baghaei@tinx.co.nz

## Abstract
Residential energy-efficiency advice is often delivered as static information that does not translate a household bill into a personalized sequence of actions. This paper presents the system architecture and operating logic of The Home-Energy Check-Up (New Zealand), a web-based public decision-support prototype designed to help New Zealand households identify avoidable energy-cost leakage, complete a short guided home inspection, and generate a prioritized behavior-first energy roadmap. The application is implemented as a single-file Python Streamlit system with session-state navigation, a household input dataclass, conservative low-high saving estimators, a seven-check inspection layer, a recommendation-ranking layer, visual analytics, anonymous Google Sheets persistence, downloadable reports, and a certificate-of-completion interface. The system does not claim to be a certified energy audit, New Zealand Building Code H1 verification method, Healthy Homes compliance statement, or guaranteed bill-forecasting engine. Instead, it operationalizes a practical educational workflow: start with money, collect only the minimum required household profile, convert user answers into a score and action set, estimate annual savings using transparent formulas, and convert behavior savings into a staged save-to-upgrade pathway. The manuscript details the front-end, state-management, calculation, data-storage, visualization, recommendation, deployment, privacy, and limitation layers of the prototype. It also identifies research-grade improvements required before the tool is used for validated impact assessment, including external validation against measured energy data, robust concurrent data writes, clearer uncertainty calibration, accessibility testing, and formal user evaluation. The contribution is a reproducible architecture for translating household energy advice into an interactive, gamified, data-light decision-support pathway for New Zealand homes.

## 1. Introduction
Residential energy-saving tools occupy a critical intersection between three distinct domains: building physics, user behavior, and communication design [1, 2]. Traditionally, the pursuit of domestic energy efficiency has been dominated by technical building science, focusing on rigorous metrics such as thermal resistance, R-values, and annual heating-load calculations [3, 4]. While these physics-grade simulations are essential for formal audits and compliance, they often create a high-friction barrier for the average occupant who may be overwhelmed by technical complexity or discouraged by the high upfront costs associated with comprehensive professional assessments [5, 6]. Consequently, a significant gap exists between the technical potential of a dwelling and the actual, everyday energy-management behaviors of its inhabitants [7, 8].
Bridging this gap requires a shift from static information delivery toward interactive decision-support systems that prioritize user engagement and energy literacy [9, 10]. Effective communication design in this context involves translating complex engineering data into accessible, actionable insights that resonate with a household's immediate concerns, financial expenditure [11, 12]. Most households do not require a full technical audit as an initial step; instead, they need a clear, low-friction diagnosis of where "bill waste" is occurring and a prioritized sequence of actions they can implement immediately. By framing the entry point around financial leakage rather than technical performance, energy-saving tools can become more accessible to non-specialist users, converting abstract concepts of building performance into a tangible "money-first" narrative.
The Home-Energy Check-Up (New Zealand), hereafter referred to as the NZ Energy Checkout prototype, addresses this first-step problem by converting a small set of user inputs into an educational, personalized action pathway. The application intentionally avoids starting with envelope modelling or compliance terminology, opting instead for a bill-framed entry point that is more intuitive for household users. By collecting the minimum required household profile and guiding users through a structured seven-check inspection, the prototype identifies avoidable energy-cost leakage and generates a staged, behavior-first energy roadmap.

The prototype is strictly positioned as an educational decision-support tool, a distinction that is vital for maintaining alignment with New Zealand's regulatory landscape. While New Zealand Building Code Clause H1 and the Healthy Homes standards provide formal compliance pathways for thermal resistance, ventilation, and moisture control, the NZ Energy Checkout prototype operates separately from these official assessment procedures. It aligns its advice with the intent of energy efficiency and healthier housing but deliberately avoids presenting its output as a certified compliance result or a guaranteed bill-forecasting engine.

Technically, the application is implemented in Python using the Streamlit framework, which supports the rapid deployment of data-driven interactive applications. The system architecture leverages st.session_state to maintain the user journey across mission screens, a HouseholdInputs dataclass to normalize user data, and a Google Sheets connection for anonymous data persistence. Interactive visualizations are handled via Plotly, while Pillow is used for the generation of completion certificates. This lightweight stack allows the system to function as a compact, reproducible web application deployable to cloud runtimes directly from a GitHub repository. This paper documents the technical architecture, algorithms, and data structures of the current prototype. It is written as a technical preprint rather than a validation paper; as such, it does not analyze measured household energy data or longitudinal behavioral outcomes. The primary contribution of this work is architectural and methodological, providing a transparent implementation model for a household-facing, money-first energy decision-support system designed specifically for the New Zealand residential context.

## 2. Scope and contribution

The prototype is not a physics-grade simulation engine. It does not perform calibrated thermal modelling, does not estimate heating degree-day demand from climate files, does not calculate code-compliant R-values, and does not claim to diagnose the actual energy performance of a particular dwelling. Instead, it performs conservative, explainable screening calculations that translate self-reported household conditions into ranked actions and estimated savings ranges. This is a useful niche because many household energy interventions fail at the first-contact stage: the user is either overwhelmed by technical information or receives generic tips that do not feel personally relevant.

The system contribution is organized around five design principles. First, the user journey is money-first: the system asks for an approximate monthly bill, household size, dwelling type, thermal condition, and main bill pain point before asking deeper questions. Second, the inspection is short and structured: users complete seven checks across living-room, bathroom, and building-shell areas. Third, estimates are ranges rather than point predictions, which is a more honest representation of uncertainty for a public prototype. Fourth, recommendations are prioritized by action relevance, tenure, bill problem, and completed checks. Fifth, the system captures only anonymous response data, while optional certificate naming is handled as an on-screen presentation feature rather than as stored research data. The layered system architecture is shown in Fig. 1. The system components that correspond to the architecture are summarized in Table 1.

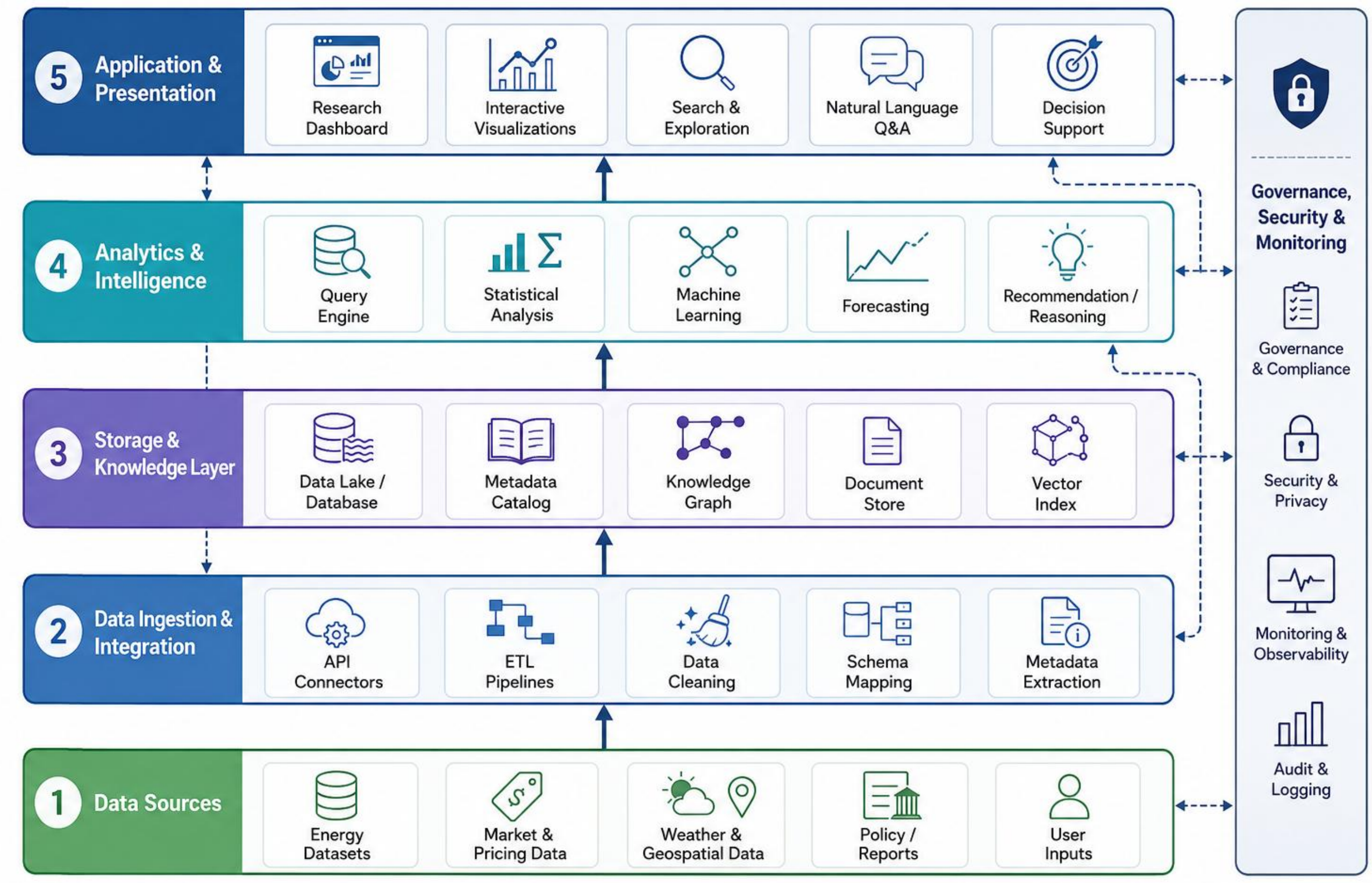

*Fig. 1. Layered architecture of the NZ Energy Checkout prototype.*

*Table 1. Main system components and responsibilities.*

| Layer | Component | Responsibility |
|---|---|---|
| Presentation | Streamlit UI | Displays mission screens, household forms, inspection checks, charts, recommendations, download controls, and certificate block. |
| State | st.session_state | Maintains stage, answers, score, selected actions, navigation history, save status, and result category across Streamlit reruns. |
| Input model | HouseholdInputs dataclass | Normalizes household size, tenure, bill problem, HVAC type, solar status, bill, tariff, shower, lighting, standby, and thermostat assumptions. |
| Decision engine | Saving and ranking functions | Computes low-high savings ranges, score labels, action priority order, top three actions, and save-to-upgrade pathway. |
| Data persistence | Google Sheets connection | Writes one anonymous row per completed participant after schema compatibility checks. |
| Visualization | Plotly and custom UI cards | Generates energy-cost pathway charts, saving bars, gauges, value cards, and recommendation visuals. |
| Output | Markdown, CSV, certificate | Allows users to download a report, export an action plan, and view a certificate of completion. |

## 3. Application context and external alignment

The prototype operates within a New Zealand residential context. Its educational advice is aligned with common household energy-efficiency behaviors and regulatory language, but the implementation remains separate from official assessment procedures. For example, the heat-pump check uses a moderate heating set-point of approximately 18-21°C, which is consistent with public EECA guidance that positions 18-21°C as a practical range balancing warmth, mold risk, and power use. The hot-water module encourages shorter showers; EECA states that reducing shower time from 15 minutes to 5 minutes can save around 66 cents per shower [13]. The lighting module prioritizes replacement of frequently used incandescent or halogen lamps with LEDs, consistent with EECA guidance that LEDs use up to 85% less electricity than traditional incandescent or halogen bulbs [14]. The standby module follows public advice to switch off unused appliances or devices on standby [15].

The tool also refers to NZ Building Code H1 and Healthy Homes thinking. H1 provides for the efficient use of energy and includes conditions affecting energy performance [16]. Healthy Homes standards establish minimum rental-property expectations in the domains of heating, insulation, ventilation, moisture ingress and drainage, and draught stopping [3]. The app uses this terminology to frame educational recommendations, especially for renters, but it repeatedly disclaims that the output is not a certified assessment.

## 4. Software stack and runtime architecture

The source file imports the Python standard library for path handling, base64 encoding, dates, in-memory bytes, text wrapping, and URL encoding. It imports pandas for tabular data handling, Pillow for image and certificate generation, Plotly graph objects for interactive figures, Streamlit for the application front end, streamlit-gsheets for Google Sheets connectivity, and dataclasses for the household input model. The stack is intentionally lightweight: all major interaction logic, estimation logic, and presentation logic are contained in one Streamlit script.

Streamlit reruns the script in response to user interaction. To maintain continuity across reruns, the application uses session state. Streamlit documentation defines session state as a mechanism for sharing variables between reruns for each user session [17]. In this prototype, session state is not a minor convenience; it is the navigation and decision backbone of the application. Variables such as current stage, score, completed hotspots, selected actions, bill risk, comfort, participant ID, response saved flag, navigation history, and the user-entered household assumptions are kept in session state.

The Google Sheets connection is created using Streamlit connection syntax. Streamlit documentation states that st.connection creates or returns a connection to a data store or API and that credentials and secrets may be supplied through Streamlit configuration such as secrets.toml [18]. In the app, the connection named gsheets is used to read existing rows, determine the next anonymous participant ID, and write updated data back to the sheet. The current implementation uses a read-modify-update pattern through the connector. This is acceptable for a prototype and small-volume deployment, but a research-grade multi-user study should use an append-safe or transaction-safe persistence method. Google Sheets API supports appending values to the next row of a detected table and updating values in a specified range; the choice of method matters for concurrency and data integrity. The user journey implemented by the runtime is shown in Fig. 2. The principal runtime technologies are listed in Table 2.

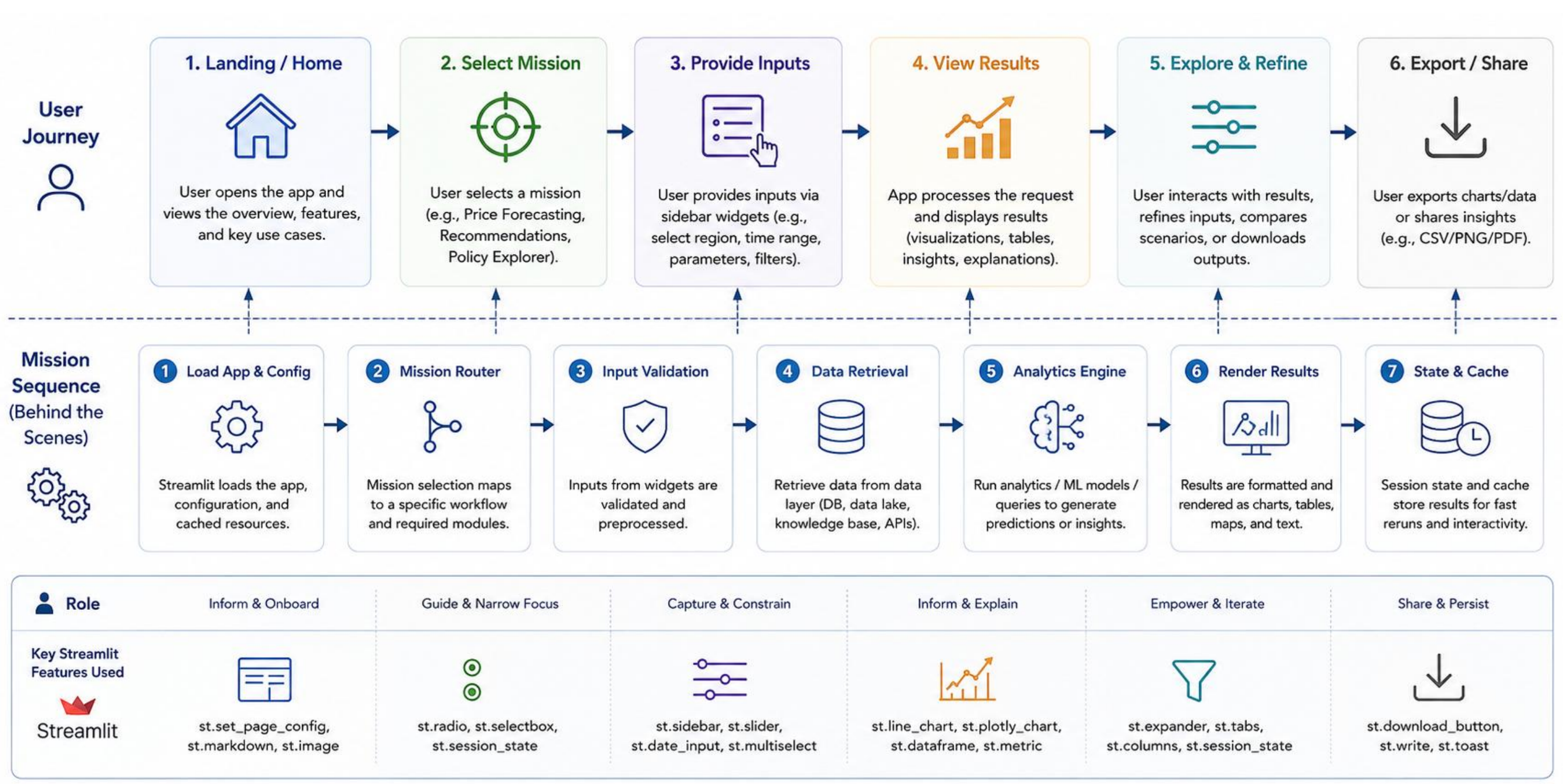

*Fig. 2. User journey and mission sequence implemented in the Streamlit app.*

*Table 2. Runtime technologies and their roles.*

| Technology | Role in prototype | Technical comment |
|---|---|---|
| Python | Core implementation language | All calculation, UI generation, persistence handling, and output generation are written in Python. |
| Streamlit | Interactive web-app framework | Provides forms, buttons, pages, reruns, sidebar, chart rendering, and downloadable outputs. |
| st.session_state | State-management layer | Stores current stage, score, answers, flags, selected actions, and navigation history. |
| pandas | Tabular data handling | Creates one-row submissions, concatenates compatible sheet rows, and builds chart/report tables. |
| Plotly graph_objects | Visual analytics | Creates line and bar figures using scatter and bar traces. Plotly graph objects support figure construction by adding traces. |
| Pillow | Certificate generation | Generates a downloadable-style PNG certificate using drawing primitives, fonts, lines, badges, and branded footer. |
| streamlit-gsheets | Google Sheets connector | Reads and updates a Google Sheet used as a lightweight anonymous response table. |
| Local assets folder | Branding and visuals | Loads company logo and optional recommendation visuals from assets/company_logo.png and related image paths. |

# 5. Data model

The central domain object is the HouseholdInputs dataclass. It provides a typed container for the variables used by the savings engine. The current fields are household_size, tenure_type, bill_problem, hvac_type, solar_status, monthly_bill_aud, electricity_price, shower_minutes_current, shower_minutes_target, showers_per_person_per_day, old_bulbs, hours_per_bulb_per_day, standby_devices, and thermostat_degrees_improved. One implementation detail should be corrected before publication-quality code release: the variable monthly_bill_aud is a legacy name from an Australian version and should be renamed to monthly_bill_nzd or monthly_bill_local for semantic accuracy. The full input model is summarized in Table 3. The data lifecycle that moves these inputs from user interaction to display and storage is shown in Fig. 3.

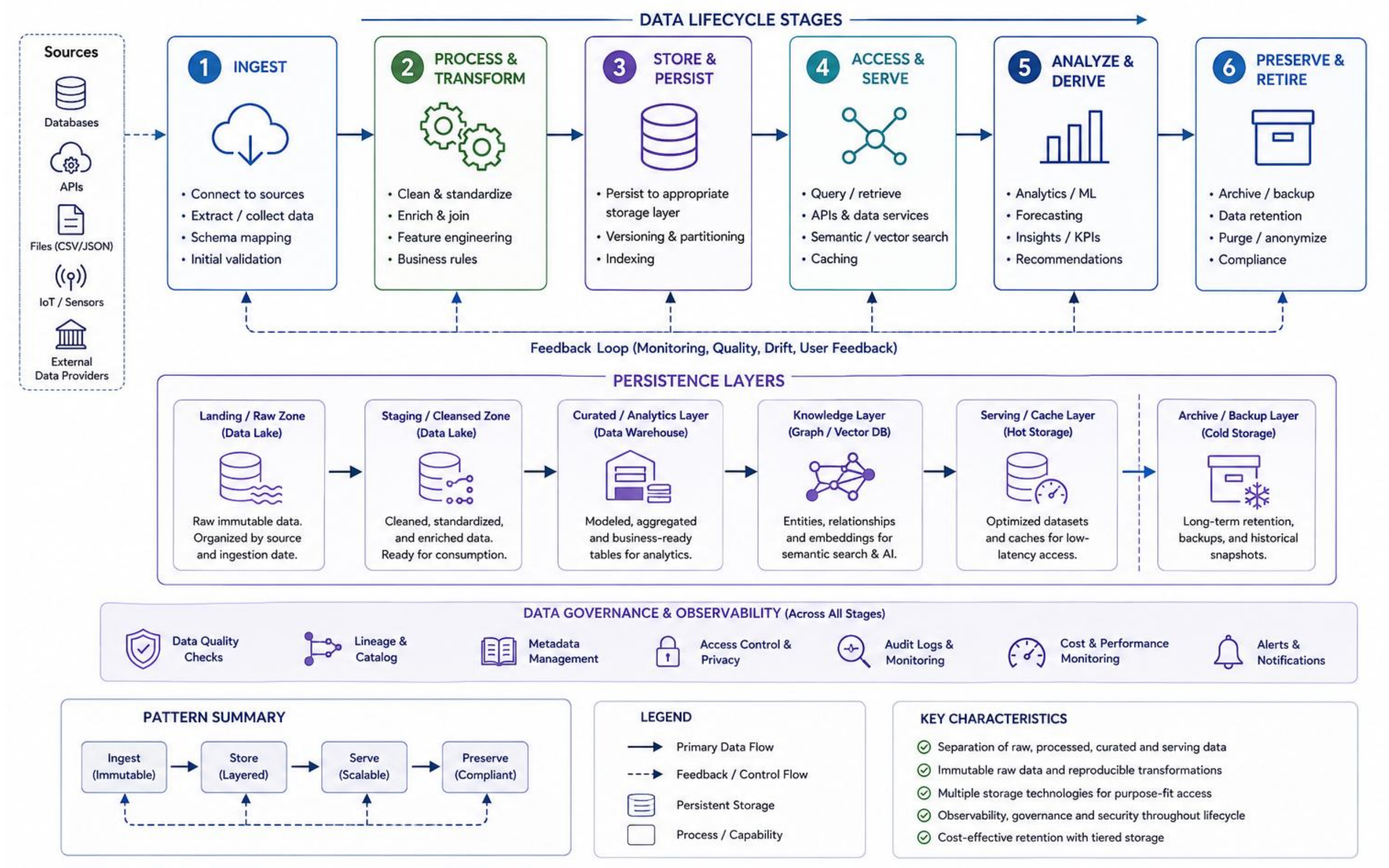

*Fig. 3. Data lifecycle and persistence pattern.*

*Table 3. Household input fields used by the calculation layer.*

| Field | Type / domain | Use in system |
|---|---|---|
| household_size | 1, 2, 3-4, 5+ | Applies people multiplier to shower hot-water savings and calibrates money-snapshot estimate. |
| tenure_type | Rented / Owned | Selects renter or owner pathway and determines whether insulation owner savings are monetized. |
| bill_problem | Winter / summer / hot water / not sure | Prioritizes actions and changes money-snapshot and thermostat estimation factors. |
| hvac_type | Heat pump, gas/fireplace, electric heater, ducted heat pump, not sure | Captured for profile and stored in response; future versions can use this for technology-specific recommendations. |
| solar_status | Yes / No / Not sure | Captured for profile and future decision rules; current ranking does not materially change by solar status. |
| monthly_bill_aud | Numeric monthly bill | Converted to annual bill. Should be renamed to monthly_bill_nzd. |
| electricity_price | NZ$/kWh | Used directly in shower, lighting, and standby saving calculations. |
| shower_minutes_current | Minutes | Subtracted from target shower duration to estimate avoidable hot-water use. |
| shower_minutes_target | Minutes | Target for the shorter-shower challenge. |
| showers_per_person_per_day | Numeric frequency | Scales hot-water calculation. |
| old_bulbs | Integer count | Scales lighting replacement saving calculation. |
| hours_per_bulb_per_day | Daily hours | Scales lighting replacement saving calculation. |
| standby_devices | Integer count | Scales standby load calculation. |
| thermostat_degrees_improved | Integer degrees | Scales thermostat set-point improvement estimate within bounded caps. |

# 6. User journey and interaction design

The application is structured as a mission-based sequence rather than a traditional audit form. The first screen frames the app as a challenge to recover money leaking from the home energy bill. The second screen asks a deliberately small set of money-first questions. The third screen collects the household profile and optional assumptions needed for saving estimates. The fourth screen guides the user through the inspection checks. The final screen generates the energy roadmap, saving estimates, confidence tags, a 30-day commitment prompt, a full recommendation list, downloadable outputs, and a certificate view.

The navigation uses a stage variable in session state. The next_stage function records the current screen in nav_history and then changes the stage. The go_back function pops from the navigation history and returns the user to the previous stage. The render_back_button function displays a back button on non-welcome pages. This approach avoids Streamlit multipage routing and keeps the prototype as a single script.

The user journey is intentionally designed to reduce friction. The money screen does not begin with technical envelope questions. Instead, it asks for the approximate bill, dwelling type, household size, perceived thermal condition, and the main bill pain point. Once the user sees a possible annual recovery range, the app asks whether they want to build a personalized saving pathway. This is a behavioral design decision: the app sells the value of continuing before asking for additional information. The screen sequence and its data purpose are summarized in Table 4.

*Table 4. Screen sequence and data purpose.*

| Screen / stage | Primary user task | System output |
|---|---|---|
| welcome | Understand mission and start challenge. | Mission framing, value proposition, and route to money scan. |
| money | Enter approximate bill and quick dwelling descriptors. | Annual spend estimate, possible recoverable waste range, monthly/three-month targets, energy-cost pathway chart. |
| profile | Add tenure, HVAC, solar, price, shower, lighting, standby, and thermostat assumptions. | Normalized HouseholdInputs object for calculation layer. |
| inspection | Complete seven checks across living-room, bathroom, and shell areas. | Score, badges, comfort/risk updates, selected actions, and eligibility to unlock roadmap. |
| plan | Review ranked actions and commit to one 30-day challenge. | Top three actions, full action list, savings chart, save-to-upgrade pathway, downloads, certificate. |

## 7. Money-first estimation module

The money-first module is a front-end engagement estimator. It is implemented by estimate_money_snapshot and returns annual_bill, waste_low, waste_high, three-month low/high, monthly low/high, dwelling_type, and home_condition. The base assumption is that a conservative avoidable-waste band exists in the entered annual energy spend. For a generic or uncertain bill problem, the base band is 8-22%. If the main problem is high winter or summer cost, the band is 10-25%. If the main problem is hot-water cost, it is 7-20%. The high end is adjusted by household size, dwelling type, and perceived thermal condition, then bounded between a minimum practical range and a maximum high-end estimate of 32%.

This module is intentionally not a measured disaggregation model. It does not separate fixed charges, network charges, fuel type, climate zone, appliance ownership, or occupancy schedules. Its purpose is to provide an interpretable, conservative opening hook. The paper should be explicit about this; otherwise, readers may incorrectly assume that the app estimates actual technical wastage. The money-first parameters are summarized in Table 5.

*Table 5. Money-first snapshot logic.*

| Decision branch | Low assumption | High assumption / adjustment |
|---|---|---|
| Default or not sure | 8% of annual bill | 22% of annual bill before calibration. |
| High winter or summer bill | 10% of annual bill | 25% of annual bill before calibration. |
| High hot-water bill | 7% of annual bill | 20% of annual bill before calibration. |
| Household size 3-4 or 5+ | No low-end change | Adds 3 percentage points to high-end estimate. |
| Detached or large detached house | No low-end change | Adds 2 percentage points to high-end estimate. |
| Small apartment / unit | No low-end change | Subtracts 2 percentage points from high-end estimate. |
| Older or draughty | Adds 2 percentage points | Adds 5 percentage points to high-end estimate. |
| Newer / efficient | Subtracts 2 percentage points | Subtracts 4 percentage points from high-end estimate. |
| Final bound | At least 4% | High estimate capped at 32% and kept above low estimate. |

## 8. Inspection logic and scoring

The inspection module is a short educational check rather than a physical audit. It uses three inspection areas: living room, bathroom, and building shell. The living-room area includes thermostat, LED, curtain, and standby checks. The bathroom area includes the shorter-shower check. The building-shell area includes draught and insulation checks. Each check asks one multiple-choice question, defines the correct answer, assigns points, and maps to a recommendation action. Correct answers increase the score, reduce bill-risk, increase comfort, and add a selected action. Incorrect answers increase bill-risk slightly and retain feedback to guide the user.

The inspection model is summarized in Table 6. The score thresholds are shown in Table 7.

*Table 6. Seven inspection checks and recommendation mapping.*

| Check | Area | Correct decision | Points | Mapped action |
|---|---|---|---|---|
| Thermostat | Living room | Heating around 18-21°C in occupied rooms. | 15 | Efficient heating/cooling set-points. |
| LED lighting | Living room / hallway | Replace frequently used old bulbs with LEDs. | 10 | LED replacement action. |
| Curtains / windows | Living room | Use curtains/blinds to block summer heat and reduce winter heat loss. | 15 | Strategic curtains, blinds, and shading. |
| Draught gaps | Door / window area | Seal obvious draughts with door snakes or weather seals. | 15 | Draught sealing. |
| Shower / hot water | Bathroom | Take shorter showers. | 15 | Shorter-shower challenge. |
| Standby appliances | Living room / study | Turn off unused devices at the wall or use a smart power board. | 10 | Standby-load control. |
| Ceiling / roof insulation | Building shell | Check or improve ceiling or roof insulation. | 20 | Insulation or renter Healthy Homes pathway. |

*Table 7. Score labels returned by the system.*

| Score range | Result label |
|---|---|
| 96-100 | Home Energy Master |
| 81-95 | Energy Efficiency Champion |
| 61-80 | Smart Saver |

| Score range | Result label |
|---|---|
| 36-60 | Home Energy Explorer |
| 0-35 | Energy Leak Beginner |

## 9. Saving-estimation engine

The core saving engine is implemented in estimate_all_savings. It takes a HouseholdInputs object and returns a dictionary of low-high annual saving estimates for shorter showers, thermostat set-point improvement, LEDs, standby control, curtains, draught sealing, owner insulation, and renter insulation. The calculations are intentionally conservative and bounded. The function itself states that the estimates are decision-support estimates and not an audit, compliance calculation, MBIE H1 verification method, Healthy Homes compliance statement, or guaranteed bill forecast.

The decision pipeline from input to roadmap is shown in Fig. 4. The formulas are summarized in Table 8. The notation uses P for electricity price, M for monthly bill, A for annual bill, H for household people multiplier, G for the positive difference between current and target shower minutes, F for showers per person per day, B for number of old bulbs, L for daily lighting hours, D for standby devices, and T for thermostat-degree improvement. The annual bill is A = 12M. The system applies a high-end cap of 45% of the entered annual bill to avoid estimates that are unrealistic relative to the household bill.

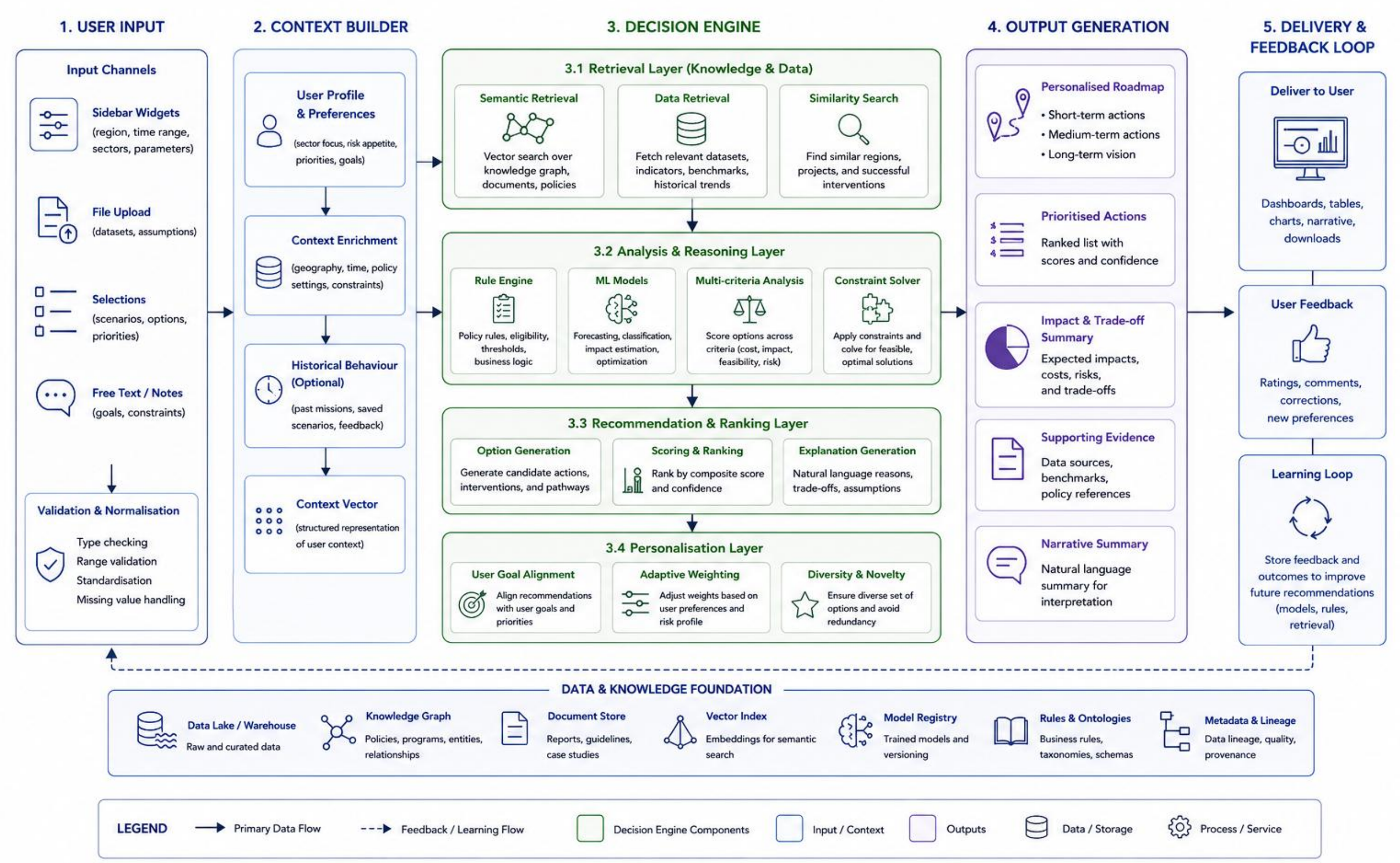

*Fig. 4. Decision engine from user input to personalized roadmap.*

*Table 8. Annual saving-estimation formulas in the current prototype.*

| Action | Low estimate | High estimate | Interpretation |
|---|---|---|---|
| Shorter showers | G x 0.30 x P x 365 x H x F | G x 0.50 x P x 365 x H x F | Assumes 0.30-0.50 kWh per avoided shower minute, scaled by household and frequency. |
| LED replacement | B x L x 365 x 0.035 x P | B x L x 365 x 0.045 x P | Assumes a conservative 35-45 W saving per old bulb replaced with LED. |
| Standby control | D x 0.003 x 24 x 365 x P | D x 0.007 x 24 x 365 x P | Assumes 3-7 W avoided standby power per controlled device. |
| Thermostat improvement | A x min(0.14, 0.025T) x factor/0.10 | A x min(0.22, 0.04T) x factor/0.10 | Scales possible bill reduction by degree improvement and whether the problem is seasonal heating/cooling. |

| Action | Low estimate | High estimate | Interpretation |
|---|---|---|---|
| Curtains and shading | NZ$60 | NZ$95 | Fixed educational estimate when annual bill is positive. |
| Draught sealing | NZ$80 | NZ$130 | Fixed educational estimate when annual bill is positive. |
| Owner insulation pathway | NZ$180 | NZ$330 | Only monetized when tenure is owned. |
| Renter insulation pathway | Not monetized | Not monetized | Framed as documentation and landlord/property-manager discussion rather than direct saving claim. |

For the default-style example of a 3-4 person rented household with a NZ$250 monthly bill, NZ$0.35/kWh electricity price, a 10-minute current shower, a 4-minute target shower, eight old bulbs used three hours per day, eight standby devices, and a two-degree thermostat improvement, the monetized total is approximately NZ$690-NZ$1,144 per year before considering non-monetized renter pathways. This example is illustrative only; the user-facing app correctly presents values as indicative ranges.

## 10. Recommendation-ranking layer

The recommendation layer is implemented through an ACTIONS dictionary, generate_ranked_actions, and top_three_actions. Each action has a priority, category, title, recommendation text, cost level, and impact level. Ranking begins with preference rules: hot-water bill problems promote shorter showers; winter and summer bill problems promote thermostat, draught, and curtain actions; tenure selects either renter or owner insulation pathway. The algorithm then appends actions completed through inspection and appends the default action sequence, removes duplicates, and sorts by priority. This creates a stable list that is both personalized and predictable.

The recommendation action library is shown in Table 9.

*Table 9. Recommendation action library.*

| Priority | Action key | Category | User-facing action | Cost level | Impact level |
|---|---|---|---|---|---|
| 1 | shorter_showers | Hot water | Run a shorter-shower challenge. | No cost | High |
| 2 | thermostat | Heating and cooling | Use efficient heating and cooling set-points. | No cost | Medium-high |
| 3 | draught_sealing | Building shell | Seal obvious draughts. | Low cost | Medium |
| 4 | curtains | Windows and comfort | Use curtains, blinds, and shading strategically. | No/low cost | Medium |
| 5 | LEDs | Lighting | Replace frequently used old bulbs with LEDs. | Low cost | Medium |
| 6 | standby | Appliances | Control standby loads. | No/low cost | Low-medium |
| 7 | insulation_owner | Building Code / retrofit | Plan insulation and envelope improvements. | Medium/high cost | High |
| 7 | insulation_renter | Healthy Homes / renting | Check rental insulation and Healthy Homes obligations. | No cost to check | Depends on landlord action |

## 11. Save-to-upgrade pathway

The system does not stop at recommending one-off tips. It constructs a funding narrative in which behavior savings can unlock later upgrades. The build_funding_pathway function first calculates the no/low-cost behavior-saving pool from shorter showers, thermostat, standby, and curtains. It then converts annual behavior savings into monthly low-high values and uses months_to_fund to estimate how long it may take for those savings to fund a next step. The pathway includes a 30-day hot-water challenge if the user has a positive shower-duration gap, LED replacement when old bulbs are reported, a conditional hot-water cylinder-wrap check when hot-water behavior is relevant, draught control, and either owner envelope-upgrade planning or renter documentation and landlord discussion.

This structure is strategically useful. The app is not merely saying “buy insulation” or “replace appliances.” It first asks the household to recover controllable cost leakage, then shows how recovered money can support incremental action. For renter households, it avoids recommending owner-controlled capital works as if they were directly under the tenant’s control. This distinction is important in New Zealand because Healthy Homes responsibilities sit with landlords for rental properties.

## 12. Visual analytics and reporting

The visualization layer serves three functions: it makes the money-first estimate tangible, supports the final roadmap, and produces shareable or downloadable outputs. The energy-cost pathway chart uses monthly seasonality arrays normalized to an annual average. Different patterns are used for high winter bills, high summer bills, high hot-water bills, and uncertain bill problems. The chart then shows three pathways: the current estimated pathway, the after-strategy pathway, and a benchmark-style standard operating band. The app explicitly states that this band is not an H1 or Healthy Homes compliance result.

The final plan screen includes value bars, a payoff strip, top three action cards, a full recommendation table, a savings bar chart, the save-to-upgrade pathway, a downloadable Markdown roadmap report, a downloadable CSV action plan, and a certificate-of-completion panel. The certificate is generated visually in the app and can also be generated by a Pillow function as a PNG. The certificate includes a disclaimer that it is a recognition certificate only, not a certified energy assessment or compliance statement.

## 13. Data persistence, privacy, and response schema

The data-persistence layer is intentionally minimal. The app saves completed anonymous responses to a connected Google Sheet. It does not collect username, email address, phone number, or home address. The participant ID is generated as "Participant N" based on the number of existing valid rows. The submission row stores household profile variables, answers and correctness flags for each check, score, result category, estimated annual savings, a string representation of the money-first snapshot, selected action titles, and consent_given=True.

The stored response schema is summarized in Table 10. The most important privacy feature is that the optional certificate display name is not included in the saved response schema. However, this does not remove all privacy considerations. Even without explicit identifiers, combinations of household size, tenure, bill level, dwelling type, and detailed responses may become sensitive if collected at scale. A public research deployment should therefore include a privacy statement, retention policy, consent language, ethics review where required, and access control for the destination Google Sheet.

*Table 10. Anonymous response schema written to Google Sheets.*

| Field group | Stored examples | Purpose |
|---|---|---|
| Participant metadata | participant_id, submission_time, consent_given | Tracks anonymous completion row and confirms the user reached the save point. |
| Household profile | household_size, tenure_type, bill_problem, dwelling_type, home_condition, hvac_type, solar_status | Supports descriptive analysis and recommendation context. |
| Inspection answers | thermostat_answer, lighting_answer, curtains_answer, draught_answer, shower_answer, standby_answer, insulation_answer | Records selected option for each check. |
| Correctness flags | thermostat_correct through insulation_correct | Allows analysis of which energy concepts users answered correctly. |
| Outcome variables | score, result_category, estimated_annual_savings, selected_actions | Stores final performance and roadmap outputs. |
| Money snapshot | money_first_snapshot | Stores the initial bill-waste estimate as a string representation of the snapshot dictionary. |

## 14. Deployment and configuration process

The deployment model assumed by the source code is a GitHub-backed Streamlit-compatible runtime. A practical deployment sequence is as follows. First, the app.py file is placed in the repository root or in the configured application path. Second, a requirements.txt file declares streamlit, pandas, plotly, pillow, streamlit-gsheets, and any connector dependencies. Third, assets/company_logo.png and any recommendation visuals are committed to the assets directory. Fourth, the deployment service is connected to the GitHub repository and configured to run the Streamlit script. Fifth, Google Sheets credentials and connection configuration are stored in the deployment platform's secret-management system, not in source code. Sixth, the Google Sheet is shared with the service account or authenticated identity used by the connector. Seventh, the public app URL is inserted into APP_PUBLIC_URL when social sharing is enabled.

The protocol-level process is simple. User interaction occurs between the browser and the Streamlit server over the deployment platform's web transport, normally HTTPS. The Streamlit server executes the Python script, stores per-user variables in session state, and renders forms, charts, and HTML/CSS. When saving occurs, the server-side connector authenticates to Google Sheets using configured credentials and writes the response table. Downloaded Markdown and CSV outputs are generated in memory and sent to the browser through Streamlit download controls.

The certificate display is rendered in the page; the Pillow function can also produce an in-memory PNG byte stream. Table 11 provides a deployment checklist for a reproducible public version.

*Table 11. Deployment checklist.*

| Item | Required configuration | Risk if missing |
|---|---|---|
| Repository | app.py, requirements.txt, assets folder, README, license. | App cannot build or lacks reproducible context. |
| Python dependencies | streamlit, pandas, plotly, pillow, streamlit-gsheets. | Runtime import failure. |
| Secrets | Google Sheets connection credentials in secrets.toml or deployment secret store. | Save function fails or credentials leak if hard-coded. |
| Google Sheet | Header-compatible sheet shared with connector identity. | Responses cannot be saved. |
| Public URL | APP_PUBLIC_URL updated from placeholder. | Social sharing links remain incorrect. |
| Disclaimers | Visible non-audit/non-compliance disclaimer retained. | Users may overinterpret outputs. |
| Monitoring | Basic logs for save errors and app exceptions. | Silent data loss or poor debugging. |

# 15. Research use and evaluation plan

The current manuscript documents the architecture but does not prove behavioral effectiveness. To turn the prototype into a publishable empirical study, a separate evaluation design is required. At minimum, the study should measure usability, comprehension, perceived relevance, intention to act, and short-term follow-up behavior. A stronger design would compare the app against a static energy-advice webpage or a PDF checklist and measure knowledge gain, action selection, and actual changes in self-reported or metered energy behavior after 4-8 weeks.

For research-grade use, the outcome variables should be aligned with the app's theory of change. The proximal outcomes are energy literacy, action prioritization, and 30-day commitment. Intermediate outcomes are self-reported implementation of selected actions and perceived bill-control self-efficacy. Distal outcomes are measured electricity reduction, reduced peak heating/cooling demand, and improved comfort. The current app already captures answers, score, selected actions, and estimated savings, but it does not yet capture baseline/post knowledge, follow-up implementation, or measured consumption. Those additions should be made before claiming behavior-change efficacy.

# 16. Limitations and technical risks

The prototype is promising, but its current state should not be oversold. The main limitation is that saving estimates are heuristic. They are transparent and conservative, but they have not yet been calibrated against measured household bills, smart-meter data, hot-water energy measurements, or thermal simulation. The second limitation is the persistence pattern. The current read-modify-update Google Sheets workflow can be vulnerable to race conditions when multiple users save at the same time. A direct append endpoint, database table, or queued serverless write function would be more robust for high-volume data collection. Third, the source file contains legacy naming from an Australian version, most notably monthly_bill_aud, which should be corrected for New Zealand publication and code release. Fourth, accessibility has not been formally audited; the app uses custom HTML/CSS cards, visual badges, and chart elements that should be tested with keyboard navigation and screen readers. Fifth, the app uses a certificate and gamified mission framing; these may increase engagement, but they should be empirically tested rather than assumed to improve learning. Table 12 lists the main limitations and recommended fixes.

*Table 12. Limitations and recommended fixes before research-grade deployment.*

| Limitation | Why it matters | Recommended fix |
|---|---|---|
| Heuristic saving estimates | Ranges may not reflect actual household consumption, fuel mix, or climate. | Validate against measured data or publish estimates as explicitly educational assumptions. |
| Read-modify-update sheet writes | Concurrent users may overwrite or lose data. | Use atomic append, a relational database, or a serverless write queue. |
| Legacy variable naming | monthly_bill_aud is semantically wrong for a New Zealand tool. | Rename to monthly_bill_nzd and update schema consistently. |
| No formal usability/accessibility evidence | A public tool must work for diverse users and devices. | Run WCAG-oriented accessibility review and usability testing. |
| No empirical outcome validation | Cannot claim energy savings or behavior change from architecture alone. | Run a controlled user study with follow-up and, ideally, energy-use data. |
| Limited privacy documentation | Anonymous data can still be sensitive in combination. | Add consent statement, retention policy, data access controls, and ethics documentation. |

| Limitation | Why it matters | Recommended fix |
|---|---|---|
| No climate-zone modelling | New Zealand heating/cooling patterns vary by region and building type. | Add region or climate input and calibrate seasonality arrays. |

## 17. Conclusion

The Home-Energy Check-Up (New Zealand) is a compact, deployable, and conceptually coherent household energy decision-support prototype. Its strongest architectural decision is the shift from generic advice to a money-first personalized pathway. The system collects minimal input, maintains state across a mission sequence, uses seven inspection checks to activate learning, computes transparent low-high savings ranges, ranks actions according to user context, and converts behavior savings into a staged save-to-upgrade roadmap. It also demonstrates a practical low-cost research infrastructure through anonymous Google Sheets persistence and downloadable outputs. The prototype should not be presented as a validated energy-saving model yet. Its scientific value at the current stage lies in architecture, workflow, and implementation logic. A rigorous next phase should correct legacy naming, strengthen data persistence, add explicit consent and privacy documentation, validate the estimation ranges, and evaluate user outcomes in a controlled study. With these improvements, the NZ Energy Checkout can become both a practical public engagement tool and a publishable platform for research on household energy literacy and behavior-first retrofit decision support.

## Data availability statement

The manuscript is based on the uploaded app.py source code for The Home-Energy Check-Up (New Zealand). For a public preprint, the author should either provide a GitHub repository URL or state that the source code will be made available upon reasonable request after intellectual-property review.

## Ethics and Privacy Statement

The current prototype stores anonymous response rows and does not require name, email address, phone number, or address. The optional certificate name is used for display and is not stored in the Google Sheets response schema. If the tool is used in a formal research study, institutional ethics review, participant information, consent language, data-retention terms, and secure data-access procedures should be added before recruitment.

## Author contribution statement

Abdollah Baghaei Daemei conceptualized the tool, prepared the prototype source code, and led the architecture and technical writing described in this manuscript.

## Conflict of interest statement

The tool is associated with Tech Innovation Experts Ltd. This association should be disclosed in any preprint or submission that discusses the system as a deployable product or service.

## References

1. Aeinehvand, R., et al., *Proposing Alternative Solutions to Enhance Natural Ventilation Rates in Residential Buildings in the Cfa Climate Zone of Rasht.* Sustainability, 2021. **13**(2): p. 679.
2. Daemei, A.B., et al., *Large-Eddy Simulation (LES) on the Square and Triangular Tall Buildings to Measure Drag Force.* Advances in Civil Engineering, 2021. **2021**(1): p. 6666895.
3. Mehrinejad Khotbehsara, E., et al., *Traditional Climate Responsible Solutions in Iranian Ancient Architecture in Humid Region.* Civil Engineering Journal, 2018. **4**(10): p. 2502-2512.
4. Baghaei Daemei, A., P. Haghgooy Osmavandani, and M. Samim Nikpey, *Study on Vernacular Architecture Patterns to Improve Natural Ventilation Estimating in Humid Subtropical Climate.* Civil Engineering Journal, 2018. **4**(9): p. 2097-2110.
5. Daemei, A.B., et al., *An experimental analysis and deep learning model to assess the cooling performance of green walls in humid climates.* Frontiers in Energy Research, 2024. **Volume 12 - 2024**.
6. Bahrami, P., et al., *Plant performance analysis of biofacades in a hot, arid Region of Qatar.* Urban Ecosystems, 2022. **25**(6): p. 1653-1678.

7. Sabouri Kenarsari, M., A. Baghaei Daemei, and M. Delshad Siyahkali, *Parametric design on daylighting and visual comfort of climatic responsive skins through Iranian traditional girih tile.* Journal of Energy Management and Technology, 2021. **5**(4): p. 29-35.
8. Baghaei Daemei, A., *Wind Tunnel Simulation on the Pedestrian Level and Investigation of Flow Characteristics Around Buildings.* Journal of Energy Management and Technology, 2019. **3**(1): p. 58-68.
9. Daemei, A.B., S.R. Eghbali, and E.M. Khotbehsara, *Bioclimatic design strategies: A guideline to enhance human thermal comfort in Cfa climate zones.* Journal of Building Engineering, 2019. **25**: p. 100758.
10. Daemei, A.B., et al., *Experimental and simulation studies on the thermal behavior of vertical greenery system for temperature mitigation in urban spaces.* Journal of Building Engineering, 2018. **20**: p. 277-284.
11. Baghaei Daemei, A., *Prototyping an AI-powered Tool for Energy Efficiency in New Zealand Homes.* ABC2: Journal of Architecture, Building, Construction, and Cities, 2025. **2025**(01): p. 20-37.
12. Daemei, A.B., R. Lovreglio, and D. Paes, *Green Roof Energy Performance across Cfb/Oceanic Climates: Simulating Climate Change and Future Trends.* Journal of Engineering, Project & Production Management, 2024. **14**(4): p. 1.
13. EECA. *Heat water efficiently at home*. 2024; Available from: https://www.eeca.govt.nz/for-homes/energy-saving-technology/water-heating/heat-water-efficiently/.
14. EECA. *Select lighting that reduces your home energy costs*. 2025; Available from: https://www.eeca.govt.nz/for-homes/energy-saving-technology/choose-good-appliances/led-lighting/.
15. EECA. *Easy ways to save on your energy bills*. 2023; Available from: https://www.eeca.govt.nz/for-homes/save-on-your-energy-bills/easy-ways-to-save-on-your-energy-bills/.
16. MBIE. *H1 Energy efficiency. Building Performance*. 2025; Available from: https://www.building.govt.nz/building-code-compliance/h-energy-efficiency/h1-energy-efficiency.
17. Streamlit. *Session State API reference*. n.d.; Available from: https://docs.streamlit.io/develop/api-reference/caching-and-state/st.session_state.
18. Streamlit. *st.connection API reference*. n.d.; Available from: https://docs.streamlit.io/develop/api-reference/connections/st.connection.